\documentclass[preprint,11pt, a4paper]{elsarticle}%

\usepackage{amssymb}

\usepackage{graphicx}
\usepackage{xcolor}
\usepackage{dirtree}

\graphicspath{{./img}}

\usepackage{float}

\restylefloat{table}

\journal{SoftwareX}

\begin{document}

\begin{frontmatter}



\title{PTRAIL - A python package for parallel trajectory data preprocessing}


\author[MUN]{Salman Haidri} 
\author[MUN]{Yaksh J. Haranwala} 
\author[UFSC]{Vania Bogorny} 
\author[CNR]{Chiara Renso} 
\author[MUN]{Vinicius Prado da Fonseca} 
\author[MUN]{Amilcar Soares}
\address[MUN]{Department of Computer Science,  Memorial University, St. John’s, NL A1B 3X5, Canada S.J. Crew Building, EN-2021}
\address[CNR]{Institute of science and technology A. Faedo, National Research Council of Italy}
\address[UFSC]{Universidade Federal de Santa Catarina (UFSC), Brazil}

\begin{abstract}

Trajectory data represent a trace of an object that changes its position in space over time.
This kind of data is complex to handle and analyze, since it is generally produced in huge quantities, often prone to errors generated by the geolocation device, human mishandling, or area coverage limitation.
Therefore, there is a need for software specifically tailored to preprocess trajectory data.
In this work we propose PTRAIL, a python package offering several trajectory preprocessing steps, including filtering, feature extraction, and interpolation.
PTRAIL uses parallel computation and vectorization, being suitable for large datasets and fast compared to other python libraries.

\end{abstract}


\end{frontmatter}

\section*{Current code version}
\label{}

\begin{table}[H]
\scriptsize
\begin{tabular}{|l|p{6.5cm}|p{7.5cm}|}
\hline
\textbf{Nr.} & \textbf{Code metadata description} & \textbf{Please fill in this column} \\
\hline
C1 & Current code version & v0.3Beta \\
\hline
C2 & Permanent link to code/repository used for this code version & $github.com/YakshHaranwala/PTRAIL$ \\
\hline
C3 & Code Ocean compute capsule & None\\
\hline
C4 & Legal Code License   & BSD 3-Clause "New" or "Revised" License \\
\hline
C5 & Code versioning system used & git \\
\hline
C6 & Software code languages, tools, and services used & python, jupyter \\
\hline
C7 & Compilation requirements, operating environments \& dependencies & numpy, pandas, geopandas, scikit-learn, hampel, scipy, psutil, folium, matplotlib, osmnx, shapely\\
\hline
C8 & If available Link to developer documentation/manual & $ptrail.readthedocs.io/en/latest/index.html$ \\
\hline
C9 & Support email for questions & amilcarsj@mun.ca \\
\hline
\end{tabular}
\caption{Code metadata (mandatory)}
\label{} 
\end{table}


\section{Motivation and significance}

Processing and extracting meaningful insights regarding the movement of people, vehicles, vessels, and animals are nowadays the focus of attention in several academic and industry sectors. 
The traces of moving objects are generally called trajectories and can be informally defined as a temporal sequence of geo-locations of a moving object.
Trajectory data often contain errors that occur due to device failure, human mistakes, or connectivity problems.
Data in this form cannot be directly used to extract useful information during data analysis. 
Besides, trajectory data in its raw format (i.e., object id, geographical position, and time) may not be sufficient for extracting relevant patterns. 
Processing trajectories using a single machine is also desired since the volume of this data is often significant. 
Indeed, there is a need for automating these preprocessing steps when dealing with trajectory data.
We call as preprocessing all steps that prepare the raw trajectory data for analysis, which in general, include 
filtering, interpolation, and feature extraction.


Currently, many python packages (e.g., scikit-mobility \cite{skmob}, as PyMove \cite{mpd}, MovingPandas \cite{pymove}) are available for trajectory data representation and manipulation. 
However, none of them focus on preprocessing trajectory data, and neither are they designed to use the available processing resources of a machine to its best.
This work introduces PTRAIL, a parallel and vectorized computational library tailored for trajectory data preprocessing. 
More specifically, PTRAIL addresses the tasks of filtering, interpolation, and feature extraction tasks performed over trajectory data. 


Trajectory data preprocessing can be a tedious and time-consuming task. 
Our objective is to allow researchers to use our library to easily handle these steps using the maximum of their resources and easily answer higher-level research questions without the need of coding and validating preprocessing steps. 
The PTRAIL functionalities and its core have been used in several projects within our group, including feature engineering in the context of transportation mode detection \cite{etemad2018predicting}, trajectory classification \cite{junior2017analytic}, and anomaly detection \cite{abreu2021trajectory}. 
However, the idea of parallezing and vectorizing the processes is the main novelty of the current version of our library.


PTRAIL has been developed in a way that it can be directly used in a Python 3 environment, and can be simply installed using the \emph{pip install} command. After installing, the user can import several functionalities available in the library and use the documentation as a reference for the requirements to use them.
First, the user should read trajectory data from a file, creating a PTRAILDataframe.
Once the data is properly loaded and validated, all the library functionalities are enabled on the data.
Furthermore, the library is open source, and therefore it can be distributed and modified by forking the repository.
The code has been well documented for the user's convenience, and several code examples are available in the developer documentation. 


The development of PTRAIL is built on top of state-of-the-art libraries that fulfilled several needs in its implementation. 
PTRAIL uses numpy \cite{harris2020array} for storing a collection of data because numpy provides a wide array of mathematical functions which are more efficient than traditional Python lists.
The PTRAILDataFrame is an extension of the pandas \cite{reback2020pandas} DataFrame and inherits all its functionalities. 
The paralellization is done using python's built in multiprocessing module \cite{multiprocessing}, which allows us to use multiple processors on a given machine by enabling concurrency and parallelism for each trajectory loaded as a PTRAILDataFrame. 
We also used geopandas \cite{kelsey_jordahl_2020_3946761} and shapely \cite{shapely} functionalities for extracting features from the data. 
We also used the implementation of Hampel filters available in \cite{hampel}, and some functionalities of scipy \cite{2020SciPy-NMeth}. 
The main point is that, in PTRAIL, the processing of the functionalities used from geopandas, shapely, Hampel filters, and scipy occur in parallel. 

\section{Software description}


PTRAIL offers several trajectory data preprocessing steps, including temporal, kinematic, and semantic feature extraction, filtering, and trajectory interpolation.
In PTRAIL we define temporal, kinematic, and semantic features as follows.
\emph{Temporal features} refers to the features that can be extracted based on the temporal dimension. Examples are the day of the week, hour, etc.
\emph{Kinematic Features} refers to those features which are concerned with the motion of the object, i.e., speed, acceleration, etc.
\emph{Semantic features} describe contextual information that is related to the movement and are generally available through geographical layers in the area where the movements occurred. Examples are the point of interest visited or the weather conditions.
The feature extraction and filtering methods allow the user flexibility to manipulate the raw data to gather information according to their application need.
Furthermore, parallel and vectorized processing has made feature extraction and filtering computationally much faster than current state-of-the-art trajectory processing libraries. 

\subsection{Software Architecture}

\begin{figure}
    \centering
    \includegraphics[width=\textwidth]{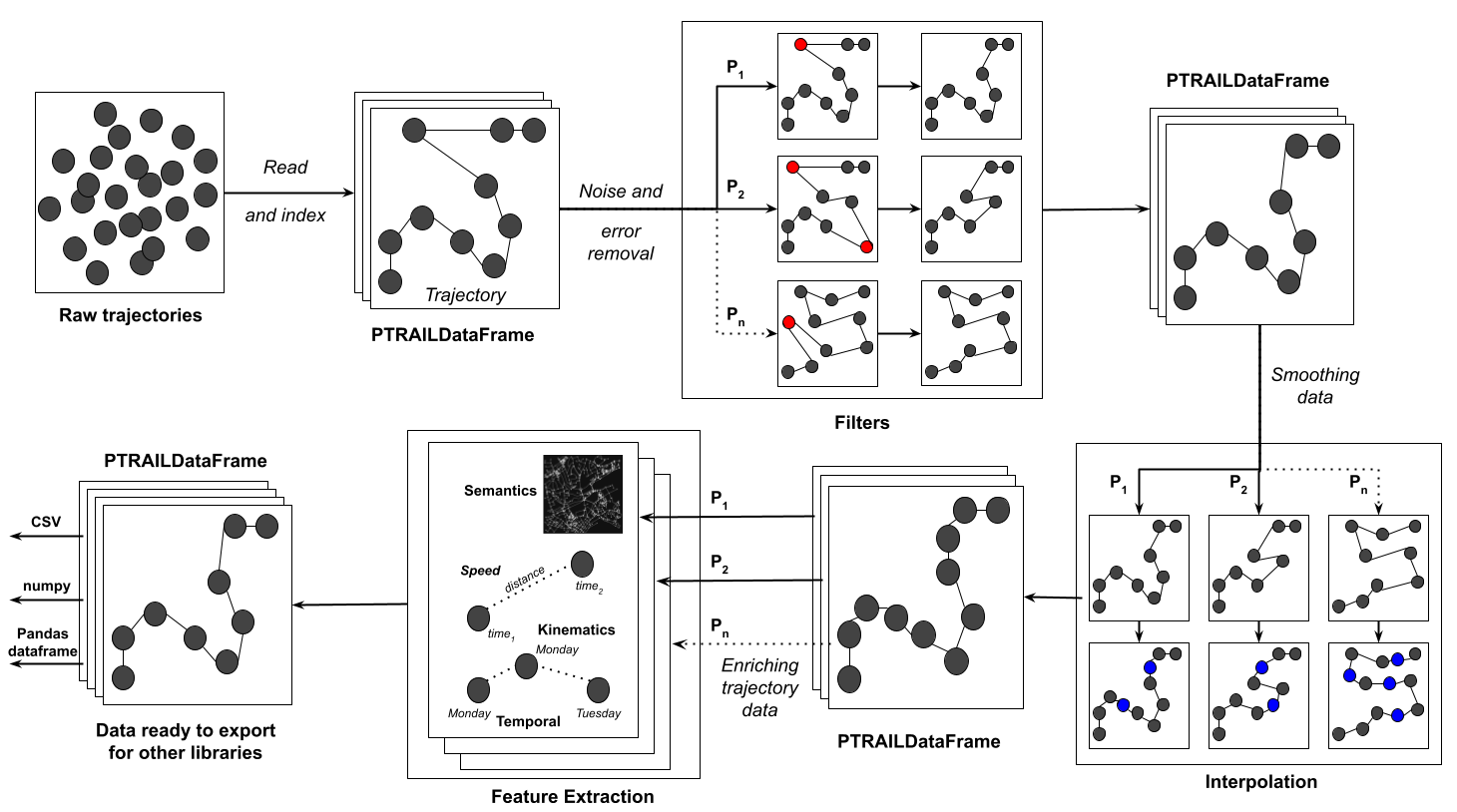}
    \caption{The PTRAIL pipeline starts by loading a PTRAIL dataframe. After, the user can apply filters for removing noise and/or smoothing the data with interpolation. Finally, features can be extracted from a higher quality data set. In the end, the data can be exported to common formats and be used with another libraries.}
    \label{fig:pipeline}
\end{figure}

The entire process of execution of a method is explained in Figure \ref{fig:pipeline}.
First, the trajectory points are fed as input for the PTRAILDataFrame. 
After loading a valid file, all functionalities are available. 
Each trajectory loaded as a PTRAILDataFrame is first vectorized and the applied functionality is executed in paralell. 
Early experiments with such a strategy showed that not overloading a machine's memory with huge matrices (expected when loading large trajectory datasets), but assembling one matrix per trajectory and paralellizing the process execution per trajectory, showed huge improvements in processing time.
The processing sequence showed in Figure \ref{fig:pipeline} is a standard way to process trajectory data. 
For each functionality, each trajectory is loaded as a matrix and the requested functionality is ran in parallel. 
The number of threads used when executing our library is a parameter that can be defined by the user. 
Indeed, it is up to the user's discretion to use filters, interpolation, and extracting useful features from the data set since the processes are independent. 

\subsection{Software Functionalities}
PTRAIL has a multitude of functionalities to be used with trajectory data, and they are detailed in the following  subsections.

\subsubsection{Filtering}

The first step when handling any raw data is to clean them  by removing noise and errors. In the trajectory data preprocessing pipeline the same step is applied. PTRAIL offers hampel, kinematic (i.e., based on speed, etc.) and temporal filters for removing errors or noise from the trajectory data sets. Our library also provides some routines to remove duplicated trajectory points since this is common in some raw data collected by geolocation devices.

\subsubsection{Interpolation}

After cleaning the data, a common next step is to smoothen it in regions where data is not available. PTRAIL offers standard ways to interpolate the data in such regions. We provide  a paralellized version of the Linear and Cubic interpolation methods from the scipy package \cite{2020SciPy-NMeth}. 
We also provide implementations of the Kinematic \cite{long2016kinematic} and Random Walk \cite{technitis2015b} strategies coded as vectors and running in parallel.

\subsubsection{Feature extraction}

Finally, after the former two steps, the user is able to extract trajectory features from the data. We provide some temporal feature extraction methods such as the extraction of the date, time, day of week from the timestamp. Such features are very useful when analyzing trajectories in contexts where analyzing repetitive actions by the moving object in particular temporal are essential, such as trajectories of people moving in cities. 
The kinematic features are essential to understand and characterize how the moving object moves over time. PTRAIL is able to calculate kinematic features such as the travelled distance, speed, acceleration, bearing, bearing rate etc. 
Finally, if geographical layers about the study area are available, the user is able to create semantic features  relating the movement with such layers. 
We provide functions that will create the moving object's visited location (inside geometry), nearest Point of Interest to a given location, functionalities to check whether trajectories lie inside a geometry or if they intersect inside one are also provided as semantic features of PTRAIL.

\section{Illustrative Examples}


\begin{figure}
    \centering
    \includegraphics[width=\textwidth]{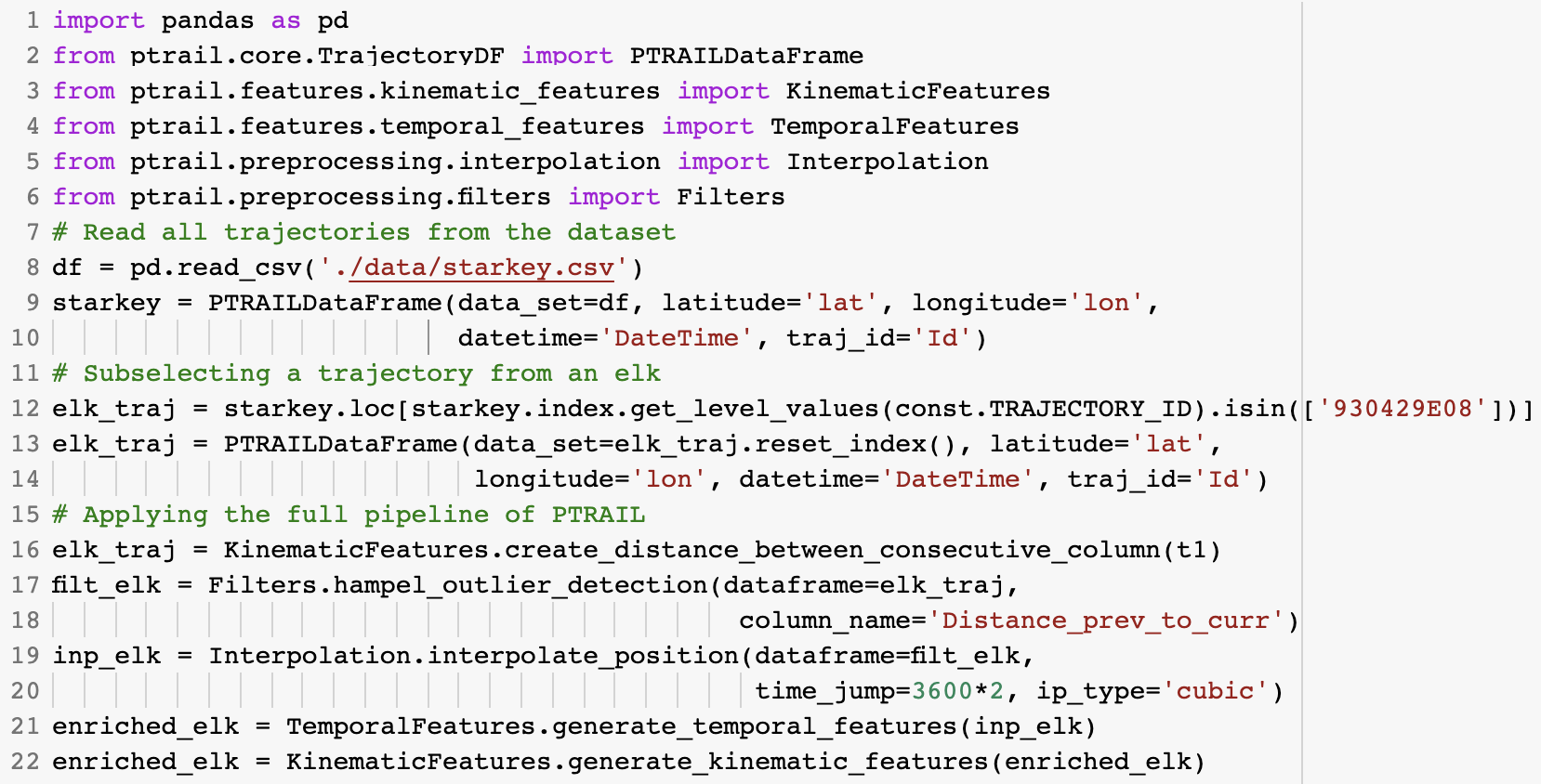}
    \caption{A code snippet for using PTRAIL with a single trajectory.}
    \label{fig:snip}
\end{figure}

\begin{figure}
    \centering
    \includegraphics[width=\textwidth]{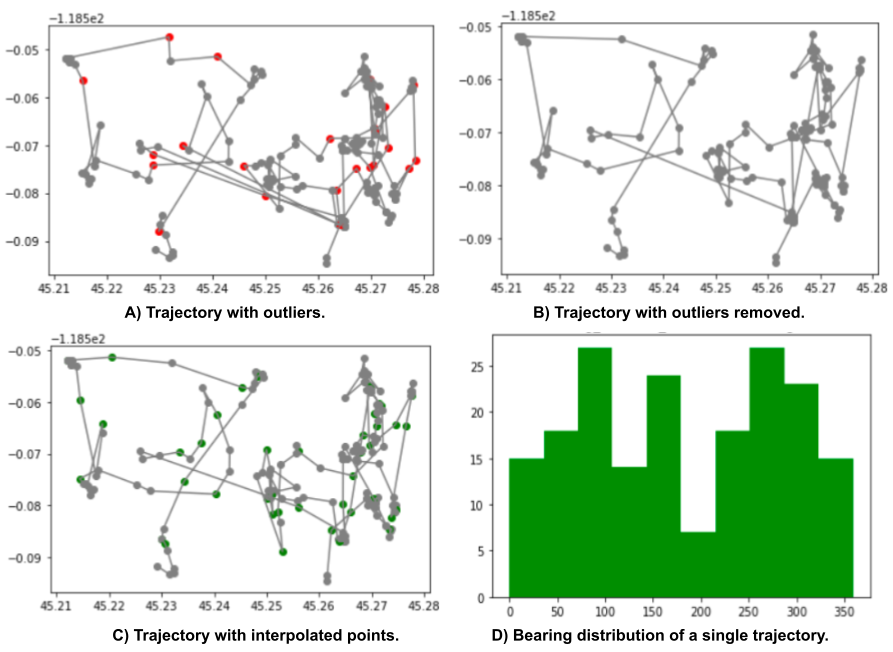}
    \caption{An illustrative example of the use of PTRAIL.}
    \label{fig:example}
\end{figure}

The PTRAIL gitHub repository currently contains numerous examples explaining the usage of PTRAIL. 
One of such examples with code snippets and outputs is illustrated in the Figure \ref{fig:example}.
Here we load a trajectory dataset with data regarding movements of cattle, elk and deer from the Starkey project \cite{starkey}. 
We load the data in lines 8-9 informing the latitute, longitude, time and trajectory id columns. 
Then, we get a single trajectory with data of an elk to pass over our the full PTRAIL pipeline (line 12). 
We decided to use a single trajectory for avoiding data cluttering of the outcomes from the PTRAIL pipeline. 
First, we create a kinematic feature with the distance between two consecutive points (line 16) since this is needed to apply a Hampel filter (lines 17-18). 
After, we interpolate the data using the cubic interpolation (lines 19-20). 
Finally, we enrich the data with all temporal and kinematic features available in the package (lines 21-22). 
Figure \ref{fig:example} shows step-by-step what happened with the data after using some of the functionalities. 
Figure \ref{fig:example}(A) shows the outliers detected by the Hampel filters and  Figure \ref{fig:example}(B) shows the trajectory with these data points removed. 
Figure \ref{fig:example}(C) shows a smoothened trajectory as the result of the cubic interpolation procedure. 
Finally, Figure \ref{fig:example}(D) shows the histogram of the bearing distribution of the elk's movement. 


\section{Impact}



PTRAIL provides a set of functionalities that can help  researchers who need to handle trajectory data for extracting information about moving objects. 
Its ability to handle noise, absence of data and data enrichment with new features in parallel makes it a convenient library for handling large trajectory datasets.
Cleaning raw trajectory data obtained from sources like Global Positioning Systems (GPS) and Automatic Identification Systems (AIS) devices is a tedious and time consuming task. 
Indeed, analysis tasks like anomaly detection, pattern mining and classification do not bring to the desired results when errors are present in the data and/or the data does not contain enough features to get interesting analysis results. 
Furthermore, the parallelization of PTRAIL's preprocessing task is particularly useful for large trajectories dataset and the ability for fast computation makes it a convenient and efficient choice for researchers.
More specifically, by using PTRAIL, researchers could focus on answering higher level research questions and focus on developing analysis methods limiting the effort into the preprocessing steps. 
Additionally, we point out that PTRAIL is organized in a modular way so that it can encourage researchers to develop more sophisticated preprocessing and semantically enriching methods that can later integrated into the library.



Since the release of its beta version in August 2021, PTRAIL surpassed 700 downloads according to PePy \cite{ptrail} and thanks to the generality of the approach (we target raw trajectories that can be collected by GPS, AIS or other location devices) we expect the usage to be spread much more. 
Plus Python is widely used by the mobility data analysis community so this library can easily integrate the stack of analysis tools available in this domain.

\section{Conclusions}
\label{}

PTRAIL is a state-of-the-art python package providing raw trajectories preprocessing tasks such as filtering, interpolation and feature engineering. 
This tool provides functions to transform raw data, as collected by location devices (therefore error prone and with noisy data) into clean and ready to use data set. 
This in turn enables the application of data mining and machine learning methods like clustering and classification. 
Furthermore, PTRAIL offers high efficiency in terms of computational speed combined with reliable and accurate results thanks to sophisticated, parallelized and vectorized calculations. 
It is also worth mentioning that PTRAIL can be suitably exploited by non expert users thanks to a large number of usage examples provided along with well documented code. 
PTRAIL follows high standards of coding and documentation which allows researchers to extend and enhance the library with newer functionalities. Finally, PTRAIL will help and reduce the researchers' burden of trajectory data preprocessing with a user friendly environment in Python for a forseeable future.

\section{Conflict of Interest}
There is no conflict of interest with this publication and there has been no significant financial support for this work that has influenced its outcome.

\section*{Acknowledgements}
\label{}

We would like to thank Memorial University of Newfounland for the Faculty of Science Undergraduate Research Award (SURA) given to Yaksh and Salman that was essential to the development of this library.  




\bibliographystyle{elsarticle-num} 
\bibliography{main_PTRAIL.bib}

\end{document}